\documentclass[5p,number]{elsarticle}

\usepackage{amssymb}
\usepackage[hyperindex,breaklinks]{hyperref}

\title{Phase imaging with intermodulation atomic force microscopy}
\author[addressKTH]{Daniel Platz}
\author[addressKTH]{Erik A. Thol\'en}
\author[addressKTH,addressSU]{Carsten Hutter}
\author[addressKTH]{Arndt C. von Bieren}
\author[addressKTH]{David B. Haviland\corref{cor1}}
\address[addressKTH]{Nanostructure Physics, Royal Institute of Technology, SE-10691 Stockholm, Sweden}
\address[addressSU]{Department of Physics, Stockholm University, SE-10691 Stockholm, Sweden}
\cortext[cor1]{Corresponding author}

\begin{document}
\begin{abstract}
Intermodulation atomic force microscopy (IMAFM) is a dynamic mode of atomic force microscopy (AFM) with two-tone excitation. The oscillating AFM cantilever in close proximity to a surface experiences the nonlinear tip-sample force which mixes the drive tones and generates new frequency components in the cantilever response known as intermodulation products (IMPs).  We present a procedure for extracting the phase at each IMP and demonstrate phase images made by recording this phase while scanning.  Amplitude and phase images at intermodulation frequencies exhibit enhanced topographic and material contrast.
\end{abstract}

\ead{haviland@kth.se}
\maketitle

\section{Intoduction}
Since its invention by Binnig et al. \cite{Binnig1986}, the atomic force microscope (AFM) has become one of the most widely used tools in nanoscience.  In the first applications of AFM only the sample topography was measured by monitoring the static deflection of the AFM cantilever. The introduction of dynamic AFM with an oscillating cantilever brought various improvements, such as much lower measurement back action which greatly reduced damage to soft samples while imaging.  In addition, the ability to measure the phase lag between the cantilever response and the exciting force, opened a new information channel to record while scanning.  These ``phase images'' enabled compositional mapping of surfaces with dynamic AFM.  The standard interpretation relates the measured phase lag to the energy dissipation resulting from the tip-sample interaction \cite{Cleveland1998}.  This interpretation is based on the assumption that the cantilever response is sinusoidal and only at the drive frequency, as in a linear approximation. 

More recently, the attention of the dynamic AFM community has focused on the nonlinear nature of the tip-sample interaction \cite{Paulo2002,Hu2006,Lee2006,Hu2008}. It has been realized that the higher harmonics of the tip-motion carry valuable information about the tip-sample interaction. Each higher harmonic represents two new information channels (amplitude and phase) that can be recorded while scanning.  Higher harmonic amplitudes and phases have been used for imaging both under ambient conditions and in liquids \cite{Hillenbrand2000,Stark2003,Preiner2007}.  With a sufficient number of these information channels, a quantitative reconstruction of the tip-sample force is possible without measuring amplitude-distance curves \cite{Stark2002,Sahin2007}. However, the signal-to-noise ratio (SNR) for higher harmonics measurements is usually very poor because the harmonics are in general not located at a cantilever resonance. Various efforts have been made to improve the signal detection, for example by modifying the cantilever design \cite{Sahin2004}.

Other approaches to increase the information acquired while scanning involve multifrequency excitations like bimodal AFM \cite{Rodriguez2004} or band excitation \cite{Jesse2007}. With bimodal AFM, two excitation tones are applied at the first two flexural eigenmodes of the cantilever. In this driving scheme the amplitude and phase at both drive frequencies are recorded as information channels. Since the feedback keeps only the amplitude at the first drive frequency constant, the oscillation at the second frequency can respond to a change of surface composition with minimal backaction \cite{Martinez2008,Lozano2008}.

Intermodulation AFM (IMAFM) \cite{Platz2008} combines the high information content of harmonics imaging with strong signal enhancement that occurs on resonance. IMAFM is inherently simpler than the aforementioned techniques because it uses only one cantilever eigenmode, making optimal use of the force sensors mechanical bandwidth. With IMAFM, a large number of new information channels are acquired with good SNR. In the following we will present a method to image surfaces by measuring the phase of the intermodulation response.

\section{Intermodulation AFM}
\begin{figure}[h]
	\includegraphics{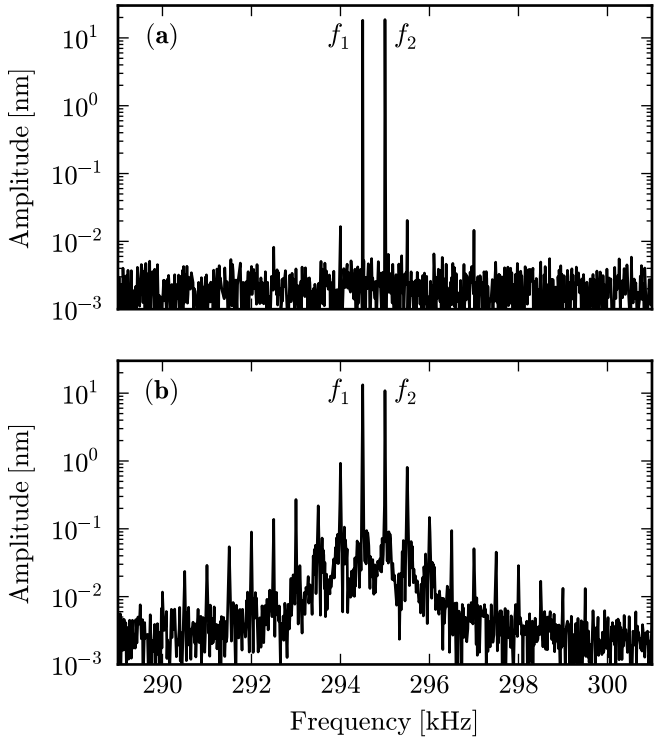}
	\caption{The measured response spectrum of a freely oscillating cantilever (a) and an oscillating cantilever which is engaging a surface(b).}
	\label{fig:spectra}
\end{figure}

The fundamental eigenmode of an AFM cantilever situated far away from a surface, and with low enough oscillation amplitude (typically less than 50 nm), can be described as a linear oscillator. When driven with two pure, harmonic drive tones $f_1$ and $f_2$ it will only show response at these two frequencies (figure \ref{fig:spectra} (a)). When the freely oscillating  cantilever is brought close to the surface, the presence of the nonlinear tip-sample force causes a frequency mixing of the drive tones (figure \ref{fig:spectra} (b)). This mixing results in a strong spectral response at frequencies $f_{\mathrm{IMP}}$ which are linear combinations of integer multiples of the drive frequencies.
\begin{equation}
f_{\mathrm{IMP}} = m\cdot f_1 + n\cdot f_2\ \ \mathrm{where}\ \ m, n \in \mathbb{Z}
\end{equation}
These new frequency components are called intermodulation products (IMPs). In figure \ref{fig:spectra} (b) the two dominant peaks in the response spectrum are located at the drive frequencies $f_1$ and $f_2$ and the other peaks in the spectrum are the different IMPs.  For this measurement, the drive tones were placed inside the first cantilever resonance. However, different driving schemes are possible where drive tones outside the resonance can be used to generate strong intermodulation response inside the resonance. The various IMPs are classified by their order $|m|+|n|$.  In figure \ref{fig:spectra} (b) several IMPs of higher order are clearly visible above the noise centered around resonance.  For this drive scheme, the IMPs appear at odd orders, $f_{3H}, f_{5H}, \cdots $ to the right of $f_2$ and similarly $f_{3L}, f_{5L}, \cdots $ to the left of $f_1$.  
 The dependence of each of the IMP amplitudes on the tip-sample separation is very complex, and when they are used for creating an image while scanning, they are very sensitive to material and chemical contrast \cite{Platz2008}.  In addition to amplitude, each IMP also has a phase and here we focus on using these phases for imaging.

\section{Experimental methods}
\begin{figure*}[h]
	\centering
		\includegraphics{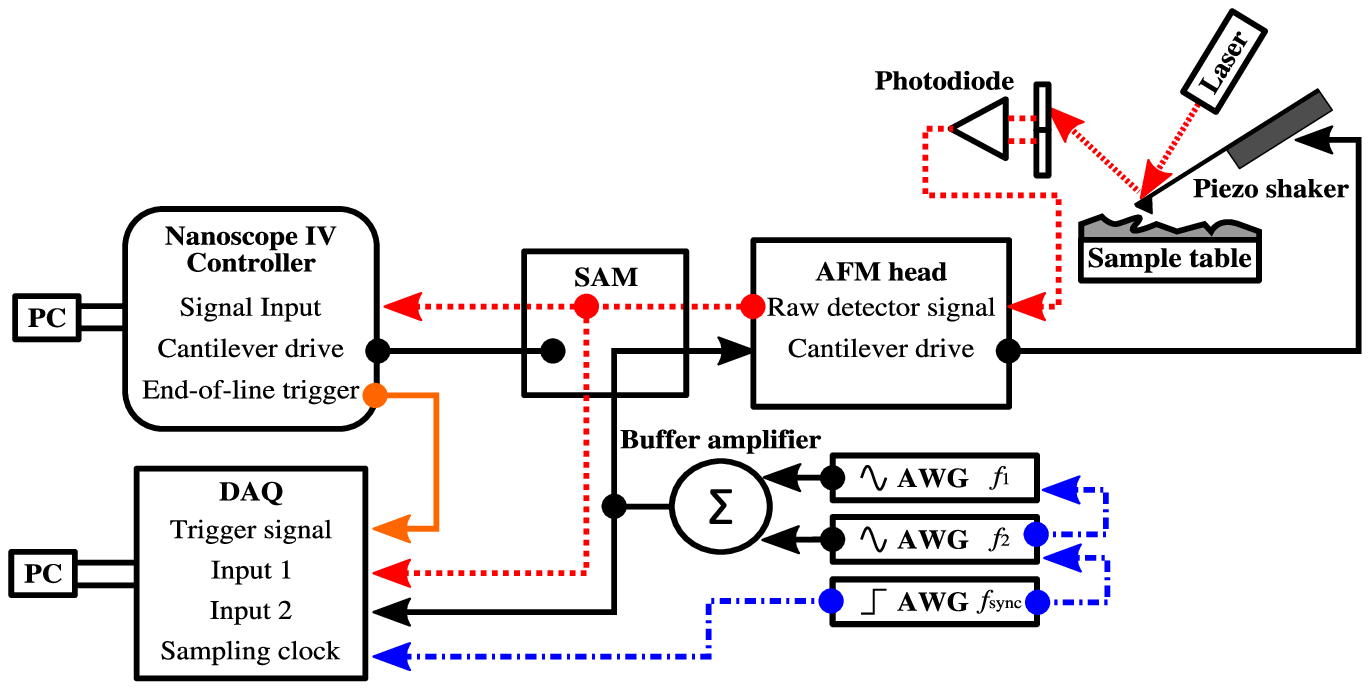}
	\caption{Experimental setup.}
	\label{fig:setup}
\end{figure*}

We use a Veeco MultiMode II and a Nanoscope IV controller (figure \ref{fig:setup}) to control the scanning and feedback while imaging. Since IMAFM requires a custom cantilever drive and data acquisition, a signal access module (SAM) is needed for IMAFM. The cantilever drive is generated by two arbitrary waveform generators (AWGs) and a buffer amplifier sums the drive tones which are applied to the piezo shaker. A data acquisition card (DAQ) measures the raw AFM photodiode signal. It is important to note that the sampling clocks of the AWGs and the DAQ have to be synchronized, and the drive and sampling frequencies must be carefully chosen in order to avoid spectral leakage at IMP frequencies which may hide the IMP response. To assure this synchronization, a third AWG generates a clock signal with frequency $f_{\mathrm{sync}}$ that controls the sampling of the photodiode signal.  $f_{\mathrm{sync}}$ can then be chosen to fit the scan speed and maximum frequency that one wishes to resolve.

Discrete Fourier transform of the photodiode signal is used to determine the amplitude and phase at the intermodulation frequencies.  The resolution bandwidth required is given by the spacing between peaks in the intermodulation spectrum, which in term sets the measurement time at each pixel of the image.  For the spectra of figure \ref{fig:spectra} we sampled for 80~ms in order to resolve the noise level between intermodulation peaks. When imaging, where speed is required, we sample for 4~ms at each pixel, which allows us to resolve intermodulation peaks spaced by 250~Hz. We convert the measured signal amplitude (volts) to cantilever deflection (nm) by calibrating the optical level sensitivity (70.4 nm/V) before imaging with a non-destructive noise calibration method \cite{Higgins2006}.

\begin{figure}[ht]
	\centering
		\includegraphics{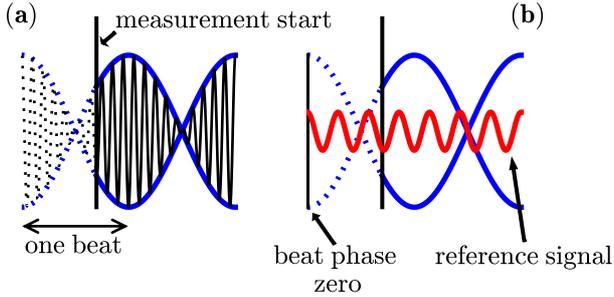}
	\caption{The phase of the beat at measurement start (a). The artificial reference signal with respect to the beat (b).}
	\label{fig:ref_phase}
\end{figure}

As one can see from the experimental setup, there is no specific reference signal available at the IMP frequencies. The determination of an IMP phase requires the computation of a reference phase that is related to the drive signals. The drive frequencies $f_1$ and $f_2$ can be chosen so that their greatest commen divisor $\Delta f$ exists. Each IMP frequency is an integer multiple of the fundamental frequency $\Delta f$.  The phase of each IMP can be related to the phase of a signal at $\Delta f$.  To see this more explicitly, we consider the case where the two drive tones are closely spaced, as in figure~\ref{fig:spectra}.  The drive tones superpose to form a beating waveform with beat frequency $\Delta f$ (figure \ref{fig:ref_phase} (a)). The phase of the beat can be determined from the phase of the two drive frequencies $f_1$ and $f_2$. We construct an artificial reference signal at each frequency $f_{\mathrm{IMP}}$, whose phase is locked to the beat (figure \ref{fig:ref_phase} (b)). The phase of the reference signal is chosen to be zero when the beat phase is zero, at the point where both drive tones are in phase and at maximum. In this way we can determine the reference phase at each pixel for an arbitrary start time of the measurement. The phase of the response at each IMP is measured relative to this reference phase.

\section{Results}
To demonstrate the ability of IMAFM to show material contrast, we fabricated a simple cross section of a stack of different materials. On a silicon dioxide ($\mathrm{SiO_2}$) substrate, layers of tantalum (Ta), magnesium (Mg), magnesium oxide (MgO) and permalloy ($\mathrm{Ni_{80}Fe_{20}}$) were deposited by sputtering. The first three layers have nominal thickness of circa 100 nm and the permalloy layer is somewhat thinner.  A focused ion beam was used to cut out a cross section, by milling at an angle of $23.5^{\circ}$ from the plane of the material stack. IMAFM scans were made on the sloped region of the stack.  

\begin{figure}[ht]
	\centering
		\includegraphics{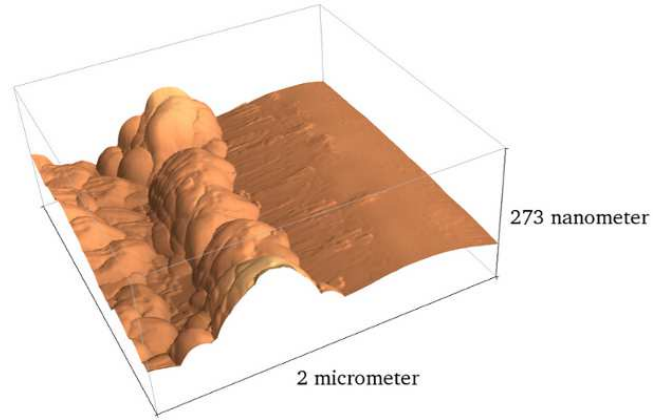}
	\caption{Sample topography}
	\label{fig:height}
\end{figure}

The topography image (figure \ref{fig:height}) shows that the initial slope is relatively smooth and a protrusion of material appears at its end. The grains in this protruded region are nearly 300 nm high and are probably caused by redeposition during ion milling, or uneven cutting due to an abrupt change in milling rate from one layer to the next.  At the far left of the image we have the $\mathrm{SiO_2}$ substrate, which also shows big height variations due to either material redeposition or inhomogeneous milling.  The height variations in the scanned area create difficulties for observing material contrast for two reasons:  The response in dynamic AFM depends on the probe-surface separation, which is not held precisely constant by the feedback when scanning over large variations in surface height.  Furthermore, the dynamic response  of the cantilever depends on the nonlinear tip-sample force, which itself is a function of the surface topography variations at the length scale of the tip radius.  Nevertheless, we are able to clearly observe layered structure in the cross-section, especially to the right of the protrusion.

\begin{figure*}[h]
	\centering
		\includegraphics{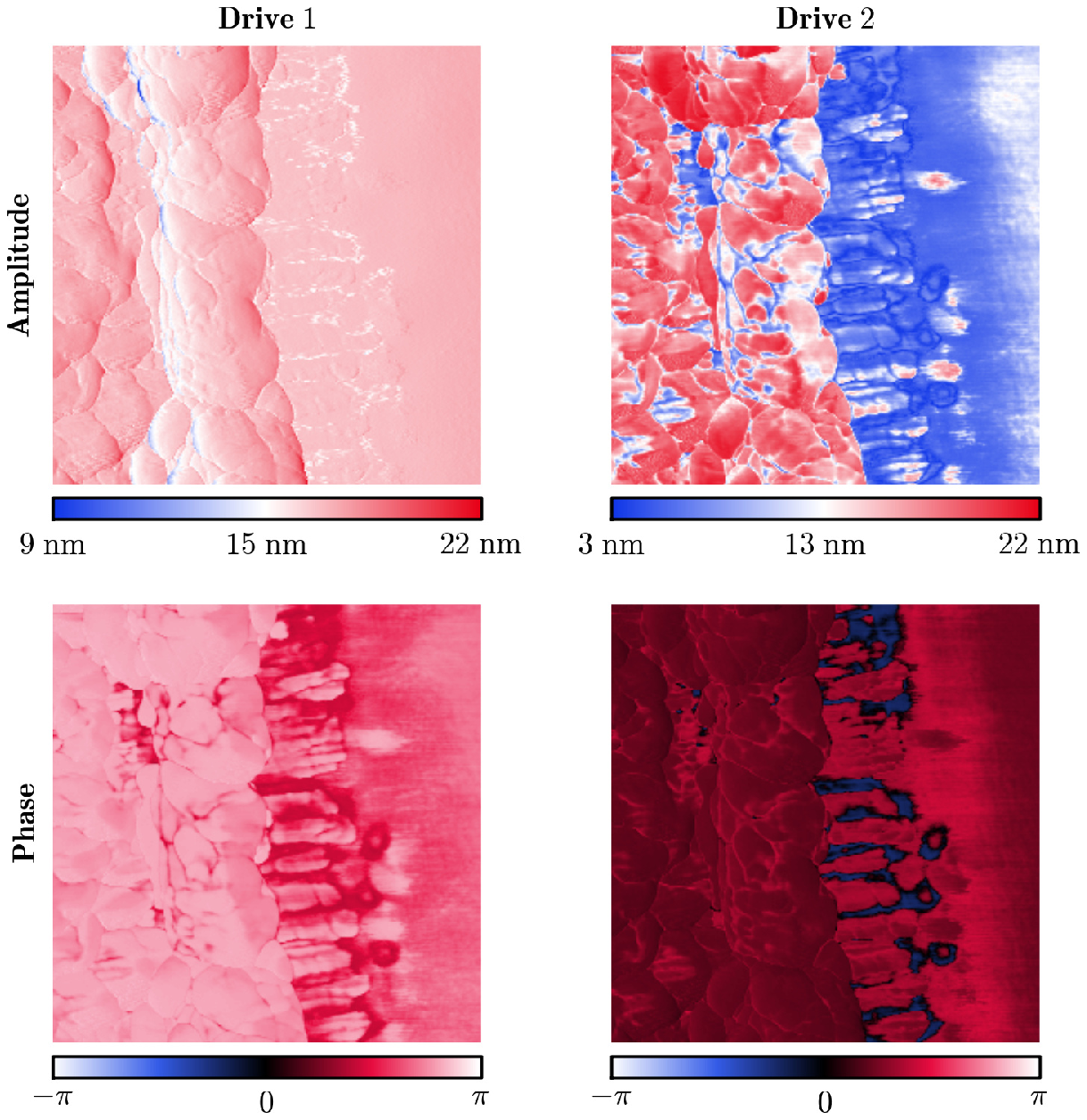}
	\caption{Amplitude and phase images at the drive frequencies $f_1$ and $f_2$.}
	\label{fig:drive}
\end{figure*}

Figure \ref{fig:drive} shows the amplitude and phase at the two drive frequencies.  The feedback was working on the amplitude of the first drive frequency $f_1$, and the features present in this image reflect the error signal, which approximately maps to the derivative of the sample topography. The phase of response at  $f_1$ (standard phase image) shows more contrast as we scan over the surface, because the feedback is not erasing this information.  Here we can see a clear contrast between the left side of the image, and the right side of the image.  On the right, we see evidence of columnar growth in one of the layers. Note that in this phase image and all following phase images, the color map is chosen so that the phases $-\pi$ and $+\pi$  are both coded with white color.  This color map reflects the fact that the phase is periodic in $2\pi$. In the image, when going from blue to red through black, there is a positive phase shift. Conversely, the phase shift is negative when going from blue to red through white. The amplitude image of $f_2$ in figure \ref{fig:drive} is not erased by the feedback, and shows more variations than the amplitude image of $f_1$. Similarly, the phase image at frequency $f_2$ also shows more variations.  Examination of these images reveals the layered structure of the stack, and substructure within the layer.  

\begin{figure*}[h]
	\centering
		\includegraphics{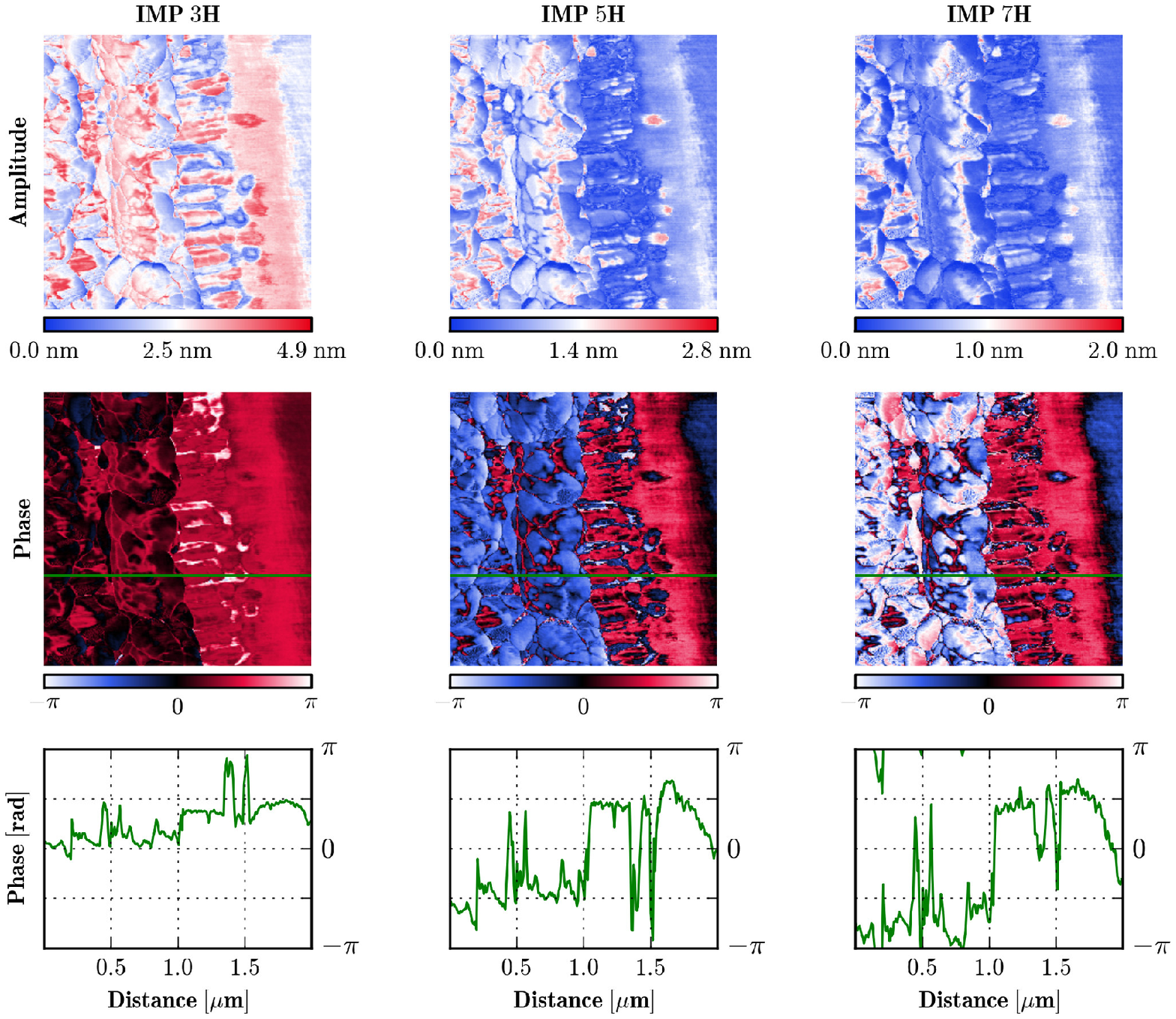}
	\caption{Amplitude and phase images at the IMP frequencies $f_{\mathrm{3H}}$, $f_{\mathrm{5H}}$ and $f_{\mathrm{7H}}$.}
	\label{fig:imps}
\end{figure*}

The amplitude and phase response at a few intermodulation frequencies is shown in figure \ref{fig:imps}. These images, as well as several similar images of higher order IMPs (now shown), were taken simultaneously with the images of figure \ref{fig:drive}, in one scan with a line scan rate of 0.5 Hz.  The IMP images have a good SNR which decreases for higher order IMP.  For the parameters chosen here ($\Delta f = 1\ \mathrm{kHz}$), we were able to clearly observe structure in images up to IMP of order 19 (i.e. at 9 times $\Delta f$ to either side of the drive frequencies), where the amplitude SNR was about 5. The features observed in figure \ref{fig:drive} are also visible in all IMP amplitude images. Especially the features induced by different materials show good contrast and we are more clearly able to see the layers of the structure.  The contrast improves with the order of the IMP, and in particular for the IMP phase, where the enhanced contrast reveals much finer structure in the image of IMP 7H than in the images of IMP 3H and IMP 5H.

In figure \ref{fig:imps} we also plot a cross section of the phase taken along the green line in each phase image.  Here we can see that the contrast, or change of phase when going from one layer to the next, is larger for higher order IMP.  Interestingly, this enhancement of the phase contrast with oder of the IMP occurs in spite of the reduction in the SNR of the IMP amplitude, which is falling off for higher order IMPs.   

\section{Summary}
We have described a method for phase measurements at IMP frequencies in IMAFM.  With this method we studied a cross section cut through layers of different hard materials.  Many images were simultaneously acquired in one scan, where both amplitude and phase of many IMPs were measured with good SNR. The IMP amplitude and phase images show significant contrast for different materials and topographic features, and in particular the phase contrast increases for higher order IMPs.  These results demonstrate that IMP imaging can be used in a similar way as higher harmonics imaging. However, in contrast to harmonic imaging, imaging with IMPs overcomes the problem of low cantilever transfer gain at the higher harmonic frequencies.  A physical model which explains in detail the origin of contrast in both amplitude and phase of the IMPs of various order, has not yet been fully illuminated, but is presently the subject of intensive research.

This work was supported by the Swedish Research Council (VR) and The Foundation for Strategic Research (SSF).

\bibliography{platz_imafm_phase_imaging}
\bibliographystyle{elsarticle-num}
\end{document}